\documentclass[10pt,twocolumn]{IEEEtran}

\usepackage{amsmath, graphics,amssymb,epsfig,subfigure,color,amsthm,cite}

\newtheorem{theorem}{Theorem}

\newtheorem{lemma}{Lemma}
\newtheorem{remark}{Remark}
\newtheorem{corollary}{Corollary}

\begin{document}

\sloppy

\title{Distributed Space-Time Interference Alignment with Moderately-Delayed CSIT}

\author{
  \IEEEauthorblockN{Namyoon Lee, Ravi Tandon, and Robert W. Heath Jr.\\}
\thanks{N. Lee and R. W. Heath Jr. are with the Wireless Networking and Communications Group, Department of Electrical and Computer Engineering, The University of Texas at
Austin, Austin, TX, 78712, USA. (e-mail:\{namyoon.lee, rheath\}@utexas.edu). The work of N. Lee and R. W. Heath Jr. was supported in part by Intel 5G program.}  
\thanks{R. Tandon is with the Discovery Analytics Center, Department of Computer Science, Virginia Tech, Blacksburg VA, 24061, USA (e-mail: tandonr@vt.edu). The work of R. Tandon was supported by the National Science Foundation under Grant CCF-1422090.}  
}

\maketitle

\begin{abstract}
This paper proposes an interference alignment method with distributed and delayed channel state information at the transmitter (CSIT) for a class of interference networks. The core idea of the proposed method is to align interference signals over time at the unintended receivers in a distributed manner. With the proposed method, achievable trade-offs between the sum of degrees of freedom (sum-DoF) and feedback delay of CSI are characterized in both the X-channel and three-user interference channel to reveal the impact on how the CSI feedback delay affects the sum-DoF of the interference networks. A major implication of derived results is that distributed and moderately-delayed CSIT is useful to strictly improve the sum-DoF over the case of no CSI at the transmitter in a certain class of interference networks. For a class of X-channels, the results show how to optimally use distributed and moderately-delayed CSIT to yield the same sum-DoF as instantaneous and global CSIT. Further, leveraging the proposed transmission method and the known outer bound results, the sum-capacity of the two-user X-channel with a particular set of channel coefficients is characterized within a constant number of bits.


\end{abstract}

\section{Introduction}
Channel state information at the transmitter (CSIT) plays an important role in interference management in wireless systems. Interference networks with global and instantaneous CSIT provide a great improvement of performance. For example, in a $K$-user interference channel with global and instantaneous CSIT, the sum degrees of freedom (sum-DoF) linearly increases with the number of user pairs $K$ \cite{Cadambe_Jafar:08}, which is much higher than that of the interference channel with no CSIT \cite{Vaze:12}. In practice, however, obtaining global and instantaneous CSIT for transmitter cooperation is especially challenging, when the transmitters are distributed and the mobility of wireless nodes increases. In an extreme case where the channel coherence time is shorter than the CSI feedback delay, it is infeasible to acquire instantaneous CSIT in wireless systems. Obtaining global knowledge of CSIT is another obstacle for realizing transmitter cooperation when the backhaul or feedback link capacity is very limited for CSIT sharing between the distributed transmitters. Therefore, in this paper, we investigate a fundamental question: is it still possible to obtain DoF benefits in interference networks under these two practical constraints$?$ In this paper we seek to answer this question by developing an interference alignment algorithm exploiting local and moderately-delayed CSIT.

Recently, an intriguing way of studying the effect of delayed CSIT in wireless networks has been initiated by the work \cite{ Maddah-Ali:12}. In particular, in the context of the vector broadcast channel,  \cite{ Maddah-Ali:12} showed that completely-delayed CSIT (i.e., CSI feedback delay larger than the channel coherence time) is still useful for improving the sum-DoF performance by proposing an innovative transmission strategy. The key idea of the transmission method was that a transmitter utilizes the outdated CSIT to align inter-user interference between the past and the currently observed signals.  
Subsequently in \cite{ Maleki:12, Tandon:13, GMK:11, AGK:13,GMK_X:11,Avestimehr,Choi:13}, the sum-DoF was investigated for a variety of interference networks (interference and X-channel) when completely outdated CSI knowledge was available at transmitters. 

Later, the characterization of the delayed CSIT effects was extended to the case where the feedback delay is less than the channel coherence time, i.e., \textit{moderately-delayed CSIT regime} \cite{Yang, Gou, Namyoon:12,Lejosne:1,Lejosne:2,Chen:1,Chen:2}. This regime is particularly interesting because, in practice, the feedback delay can be less than the channel coherence time depending on the user mobility. Further, by leveraging the results obtained with completely-delayed CSIT, it is possible to obtain a complete picture on how the CSI feedback delay affects the scale of the capacity. In the moderately-delayed CSIT regime, a transmitter is able to exploit alternatively both current and delayed CSI. This observation naturally leads to the question of whether current and delayed CSIT can be jointly exploited to obtain synergistic benefits. In recent work in \cite{Namyoon:13,Tandon_alt:13}, the benefits of jointly exploiting current and outdated CSIT are substantial over separately using them in the context of vector broadcast channel. In particular, it was shown that, up to a certain delay in the feedback, sum-DoF loss does not occur for the multi-input-single-output (MISO) broadcast and interference channel \cite{Namyoon:12, Namyoon:13,Namyoon_IC:12}.


Local CSIT is also a preferable requirement in the design of wireless systems particularly when the transmitters are not co-located, and the capacity of the backhaul links is limited. When the transmitters are distributed, each transmitter may obtain local CSI between itself and its associated receivers using feedback links without further exchange of information between the transmitters. The impact of local or incomplete CSIT has been actively studied, especially for multiple-input-multi-output (MIMO) interference networks \cite{Gomadam:11, Berry:09, Peter:11, Gesbert_networkMIMO:12, Gesbert:12,  GMK:11}. In particular,  \cite{Gomadam:11, Berry:09, Peter:11, Gesbert_networkMIMO:12} proposed iterative algorithms for interference alignment with local but instantaneous CSIT and demonstrated the DoF benefits in MIMO interference channels. The limitation, however, is that the convergence of the algorithms in  \cite{Gomadam:11, Berry:09, Peter:11} is not guaranteed when CSI delay is considered. In \cite{Gesbert_IA:12, Gesbert:12}, the feasibility of interference alignment was characterized by an iterative algorithm using incomplete but instantaneous CSIT in a $K$-user MIMO interference channel. In \cite{GMK:11, GMK_X:11} it was shown that it is possible to strictly increase the DoF with completely-delayed and local CSIT for the the two-user MIMO and $K$-user MISO interference channels which are more closely related to our work. Nevertheless, to the best of our knowledge, characterizing the benefits of DoF is still an open problem in single-antenna interference channels with local and moderately-delayed CSIT.

In this paper we propose a distributed interference management technique for interference networks with distributed and moderately-delayed CSIT.
The proposed method is a structured space-time repetition transmission technique that exploits both current and outdated CSIT jointly to align interference signals at unintended receivers in a distributed manner. 
Since this method is a generalization of the space-time interference alignment (STIA) in a vector broadcast channel \cite{Namyoon:12, Namyoon:13} by imposing the distributed CSIT constraint, we refer to it as ``\textit{distributed STIA}." One distinguishing feature of the proposed method is that a transmitter only uses local CSIT, thereby reducing overhead incurred by CSIT sharing between transmitters, which differs from the conventional IA in \cite{Cadambe_Jafar:08} and STIA in \cite{Namyoon:12, Namyoon:13}. With the proposed method, we show that the optimal sum-DoF $\frac{2K}{K+1}$ is achievable for the $K\times 2$ X-channel with local CSIT, provided that the CSI feedback delay is less than $\frac{2}{K+1}$ fraction of the channel coherence time. This result implies that there is no loss in DoF even if the transmitters have local and delayed CSIT for this class of X-channels. Furthermore, for the three-user interference channel with local CSIT, we demonstrate that a total of $\frac{6}{5}$ sum-DoF is achievable when the CSI feedback delay is three-fifths of the channel coherence time, i.e., $T_{\rm fb}\leq \frac{3}{5}T_{\rm c}$. By leveraging the sum-DoF result in \cite{AGK:13} with our achievability results, we establish inner bounds of the trade-off for both channels. A major implication of this result is that local and moderately-delayed CSIT obtains strictly better the sum-DoF over the no CSIT case in a certain class of interference channels. As a byproduct, leveraging the sum rate outer bound result in \cite{Cadambe_Jafar:09} and the proposed method, we characterize the sum-capacity of the two-user X-channel with a set of particular channel coefficients within a constant number of bits.

The rest of the paper is organized as follows. In Section II, we describe signal models of the X-channel, the three-user interference channel, and CSI feedback model. We explain the key idea of the proposed transmission method through a motivating example in Section III.  In Section IV,  we characterize the trade-off region between the sum-DoF and CSI feedback delay of the X-channel. We provide an inner bound of the trade-off region for the three-user interference in Section V.  In Section, we provide the sum-capacity results for the two-user X-channel within a constant number of bits. The paper is concluded in Section VII. 

Throughout this paper, transpose, conjugate transpose, and inverse of a matrix ${\bf A}$ are represented by ${\bf A}^{T}$, ${\bf A}^{*}$, ${\bf A}^{-1}$, respectively. In addition, $\mathcal{CN}(0,1)$
represents a complex Gaussian random variable with zero mean and unit variance. 

\section{System Model}
In this section, we explain two signal models for the $K\times 2$ X and three-user interference channel and describe the CSI feedback assumptions used for this paper. 
\begin{figure}
\centering
\includegraphics[width=3.0in]{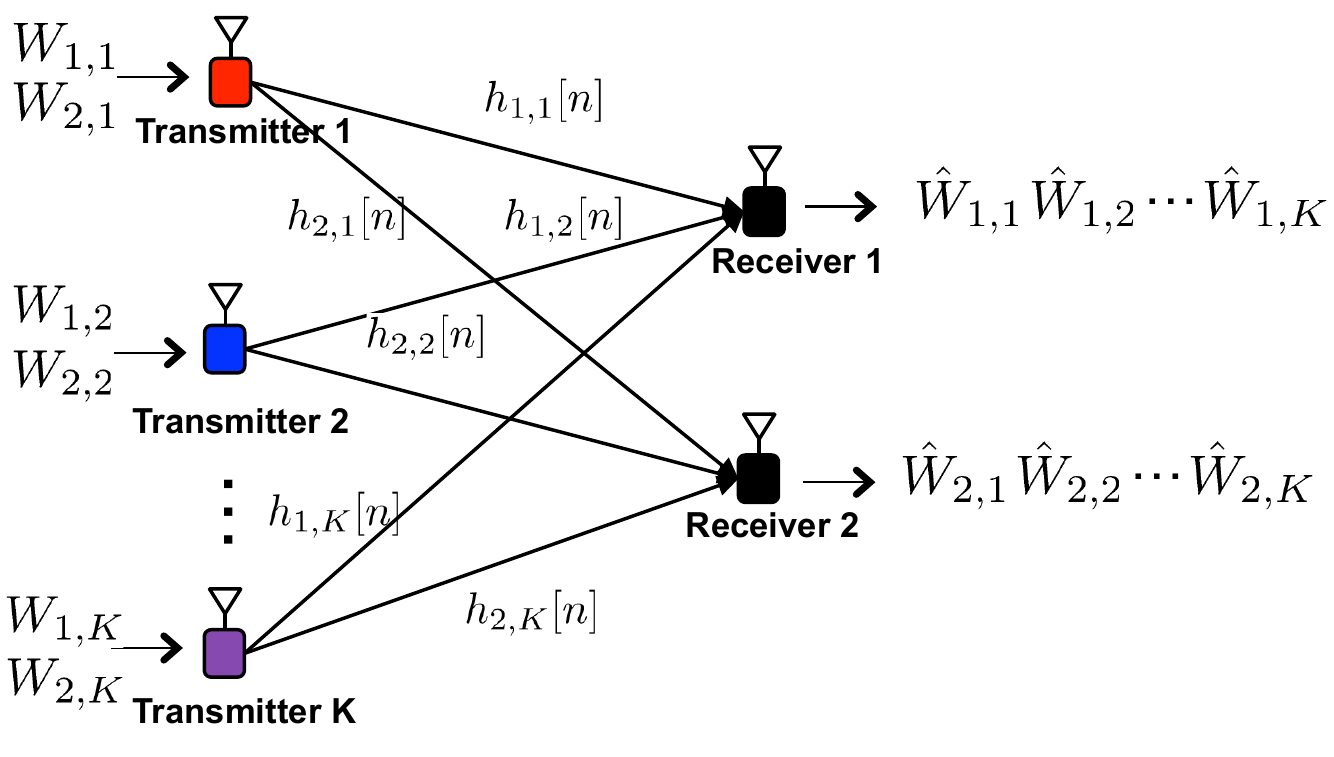}
\caption{Illustration of the $K\times 2$ X-channel with a single antenna. } \label{fig:1}\vspace{-0.1cm}
\end{figure}

\subsection{Signal Models}

\subsubsection{$K\times 2$ X-channel} 

We consider the X-channel with $K$ transmitters and two receivers, each with a single antenna. As illustrated in Fig. \ref{fig:1}, transmitter $k\in\{1,2,\ldots,K\}$ intends to send an independent message $W_{\ell,k}$ for receiver $\ell\in{1,2}$ using input signal $x_k[n]$. When the $K$ transmitters simultaneously send their signals in time slot $n$, the received signal $y_{\ell}[n] \in \mathbb{C} $ at receiver $\ell\in\{1,2\}$ is
\begin{align}
{y}_{\ell}[n] &= \sum_{k=1}^K{h}_{\ell,k}[n]x_{k}[n] + {z}_{\ell}[n],
\end{align}
where ${z}_{\ell}[n]$ denotes the additive noise signal at receiver $\ell$ in time slot $n$ which elements
are Gaussian random variable with zero mean and unit variance,
i.e., $\mathcal{CN}(0,\sigma^2)$, and ${{h}_{\ell,k}}[n]$ represents the channel value from transmitter $k$ to user $\ell$. All channel values in a different fading block are drawn from an independent and identically distributed (IID) continuous distribution across users. 

Assuming that feedback links are error-free but have the feedback delay of $T_{\textrm{fb}}$ time slots, transmitter $k\in\{1,2,\ldots,K\}$ has knowledge of the channel vector ${\bf h}_{\ell,k}^{n-T_{\rm fb}}=\{h_{\ell,k}[1],h_{\ell,k}[2],\ldots, h_{\ell,k}[n-T_{\rm fb}]\}$ up to time $n$ for two receivers $\ell\in\{1,2\}$. We denote the local and delayed CSI matrix known to transmitter $k$ in time slot $n$ by ${\bf H}_{k}^{n-T_{\rm fb}}=\left[{\bf h}_{1,k}^{n-T_{\rm fb}}, {\bf h}_{2,k}^{n-T_{\rm fb}}\right]$. Then, the input signal of transmitter $k$ is generated as a function of the transmit messages and the delayed and local CSIT, i.e., $x_k[n]=f_k(W_{1,k},W_{2,k}, {\bf H}_{k}^{n-T_{\rm fb}})$ where $f_k(\cdot)$ represents the encoding function used by transmitter $k$. The transmit power at each transmitter is assumed to be $P$, $\mathbb{E}\left[|{x}_{k}[n]|^2\right] \leq P$.

\subsubsection{ Three-User Interference Channel} 
\begin{figure}
\centering
\includegraphics[width=3.0in]{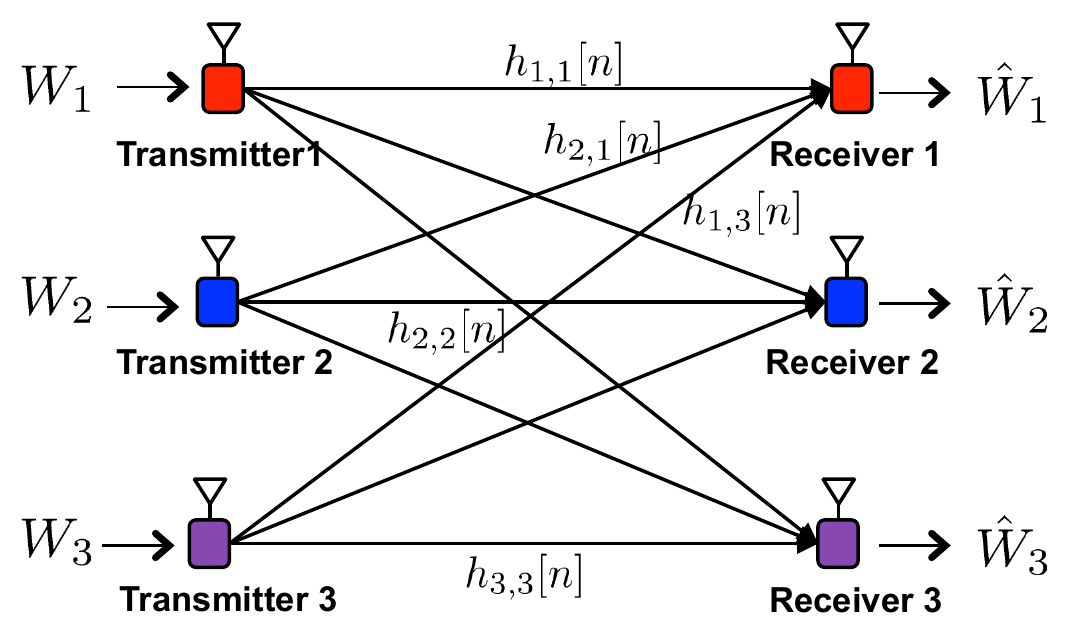}
\caption{Illustration of the three-user interference channel with a single antenna.} \label{fig:2}
\end{figure}

We also consider the three-user interference channel where all the transmitters and the receivers have a single antenna as illustrated in Fig. \ref{fig:2}. The difference with the X-channel is that, in this channel, transmitter $k$ for $k\in\{1,2,3\}$ intends to send one unicast message $W_k$ to its corresponding receiver $k$ using input signal $x_k[n]$. 
Denoting the local and delayed CSI matrix known to transmitter $k$ up to time slot $n$ by ${\bf H}_{k}^{n-T_{\rm fb}}=\left[{\bf h}_{1,k}^{n-T_{\rm fb}}, {\bf h}_{2,k}^{n-T_{\rm fb}}, {\bf h}_{3,k}^{n-T_{\rm fb}}\right]$, the input signal is generated by a function of the message $W_k$ and the channel knowledge ${\bf H}_{k}^{n-T_{\rm fb}}$, i.e., $x_k[n]=f_k(W_{1,k}, {\bf H}_{k}^{n-T_{\rm fb}})$. Then, the channel output at receiver $\ell$, $y_{\ell}[n]$, is given by
\begin{align}
y_{\ell}[n]=\sum_{k=1}^3h_{\ell,k}[n]x_k[n]+z_{\ell}[n],
\end{align}
where the transmit power at each transmitter is assumed to be $P$, $\mathbb{E}\left[|{x}_{k}[n]|^2\right] \leq P$.

%

\subsection{Block Fading and CSI Feedback Model}

\begin{figure}
\centering
\includegraphics[width=3.2in]{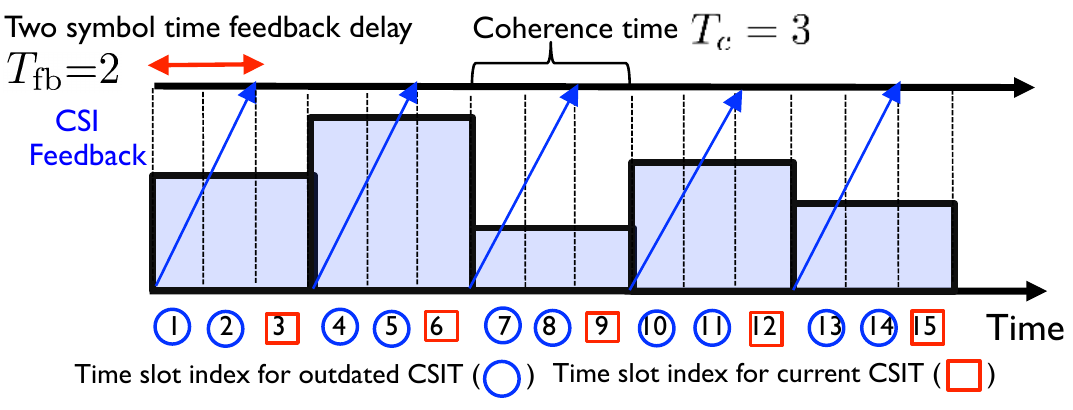}
\caption{When the channel coherence time is $T_{\rm c}=3$ and CSI feedback delay is $T_{\rm fb}=2$, a transmitter is able to access current and outdated CSI over one-thirds and two-thirds of the channel coherence time. } \vspace{-0.1cm}\label{fig:3}
\end{figure}

We consider ideal block fading channels where the channel realization remains constant within a block of certain length, i.e., channel values are invariant for the channel coherence time $T_{\textrm{c}}$. Each block is an independent realization. As illustrated in Fig. \ref{fig:3}, each receiver perfectly estimates CSI from different transmitters and sends back it to the corresponding transmitters every $T_{\textrm{c}}$ time slots periodically through error-free but delayed feedback links. This periodic CSI feedback model allows for the transmitters to continuously track all variations in the channel changes. It is worth to note that our block fading assumption differs from the time-correlated block fading models used in \cite{Yang, Gou,Chen:1,Chen:2}. Since the channel values are assumed to be changed within the same channel coherence block in the correlated model, the transmitter is able to partially track variations in the channel changes, thereby providing imperfect current CSIT.

We assume that the feedback delay time $T_{\textrm{fb}} $ is less than the channel coherence time, $T_{\textrm{fb}}<T_{\textrm{c}}$ as in \cite{Namyoon:12, Namyoon:13}. If a receiver sends back CSI in time slot $n$, the transmitter is able to use the CSI from time slot $n+T_{\textrm{fb}}$. Since the channel variation is slower than the feedback speed $T_{\textrm{fb}}<T_{\textrm{c}}$, transmitter $k$ has the set of current and delayed channel knowledge ${\bf h}_{\ell,k}^{n-T_{\rm fb}}=\{h_{\ell,k}[1],h_{\ell,k}[2],\ldots, h_{\ell,k}[n-T_{\rm fb}]\}$ in time slot $n$. For instance, as depicted in Fig. \ref{fig:3}, in time slot 9, the transmitter has delayed channel knowledge for the first and second blocks, whereas it has current channel knowledge for the third block.

Let us introduce a parameter that measures CSI obsoleteness compared to the channel coherence time. We refer to this parameter as the normalized CSI feedback delay:
\begin{align}
\lambda=\frac{T_{\textrm{fb}}}{T_{\textrm{c}}}.
\end{align}
The case of $\lambda\geq 1$ corresponds to the case of completely outdated CSI regime as considered in \cite{Maddah-Ali:12}. In this case, only completely outdated CSI  is available at the transmitter. We refer to the case where $\lambda=0$ as the instantaneous CSIT point. Since there is no CSI feedback delay, the transmitter can use the current CSI in each slot. As illustrated in Fig. \ref{fig:3}, if $\lambda=\frac{2}{3}$, the transmitter has instantaneous CSI over one-thirds of the channel coherence time and completely outdated CSI for the previous channel blocks.

\subsection{Sum-DoF and CSI Feedback Delay Trade-Off}

Since the achievable data rate of the users depends on the parameters $\lambda$ and signal-to-noise ratio (SNR), we express it as a function of $\lambda$ and $\textrm{SNR}$. In particular, for codewords spanning $n$ channel uses, a rate of message $W_{\ell,k}$,
$R_{\ell,k}(\lambda,\textrm{SNR})=\frac{\log_2|W_{\ell,k}(\lambda,\textrm{SNR})|}{n}$, is achievable if the
probability of error for the message $W_{\ell,k}$ approaches zero as
$n\rightarrow \infty$. Then, the sum-DoF trade-off of the X-channel with local CSIT is defined as a function of the normalized feedback delay,
\begin{align}
d^{\rm X_{L}}_{\Sigma}(K,2;\lambda) =\lim_{\textrm{SNR}\rightarrow \infty}
\frac{\sum_{\ell=1}^2\sum_{k=1}^{K}R_{\ell,k}(\lambda,\textrm{SNR})}{\log(\textrm{SNR})}.
\end{align}
 With the same definition of a rate for message $W_k$, the sum-DoF trade-off of the three-user interference channel with local CSIT is
\begin{align}
d^{\rm IC_{L}}_{\Sigma}(3,3;\lambda) =\lim_{\textrm{SNR}\rightarrow \infty}
\frac{ \sum_{k=1}^{3}R_{k}(\lambda,\textrm{SNR})}{\log(\textrm{SNR})}.
\end{align}
Note that the sum-DoF regions of interference networks with local and delayed CSIT are less or equal to those of the networks with global and delayed CSIT.

\section{Proposed Transmission Method}
In this section, we illustrate the core ideas behind our approach using the $2\times 2$ X-channel as an example. Gaining insights from this section, we extend our method into the $K\times 2$ X-channel and three-user interference channel in a sequel.
 
 Let us consider the $2\times 2$ X-channel with local and delayed CSIT and focus on the case of $\lambda=2/3$, i.e., when feedback time is two-thirds of the channel coherence time. It is known from \cite{MMK:08}, that the sum DoF with instantaneous CSIT (corresponding to $\lambda=0$ is 4/3). We will show that even with $\lambda=2/3$, the sum DoF of $4/3$ is achievable. In particular, we will show that four independent data symbols are delivered over 3 channel uses $\{h_{i,j}[1],h_{i,j}[4],h_{i,j}[9]\}$. In time slot 9, a transmitter with the set of delayed CSI $\{h_{i,j}[1],\ldots,h_{i,j}[7]\}$ is able to access current CSI  because $h_{i,j}[7]=h_{i,j}[9]$ as depicted in Fig. \ref{fig:3}. Note that since the all three time slots belong to different channel coherence blocks, all elements of the channel are IID random variables.
 
The proposed transmission method involves two phases.
\textbf{Phase One:} Phase one spans two time slots. In the first time slot, transmitter 1 and transmitter 2 send information symbols $x_1[1]=a_1$ and $x_2[1]=b_1$, which are intended for receiver 1. Ignoring noise, receiver 1 obtains a linear combination of the desired symbols, while receiver 2 overhears one equation as
\begin{align}
{ y}_{1}[1] &={h}_{1,1}[1]a_1+  {h}_{2,2}[1]b_1=L_1[1](a_1,b_1),  \\
{ y}_{2}[1] &={h}_{2,1}[1]a_1+  {h}_{2,2}[1]b_1=L_2[1](a_1,b_1).
\end{align} 
In time slot 4, transmitter 1 and transmitter 2 send signals $x_{1}[4]=a_2$ and $x_{2}[4]=b_2$ for receiver 1. Then, receiver 2 obtains a linear combination of desired symbols while receiver 1 listens a linear equation for unintended symbols as
\begin{align}
{ y}_{1}[4] &={h}_{1,1}[4]a_2+  {h}_{2,2}[4]b_2=L_1[4](a_2,b_2), \\
{ y}_{2}[4] &= {h}_{2,1}[4]a_2+  {h}_{2,2}[4]b_2=L_2[4](b_2,b_2).
\end{align}

\textbf{Phase Two:} In time slot 9, transmitter 1 and 2 are able to access local and delayed CSIT, i.e., $\{h_{1,1}[n],h_{2,1}[n]\}$ and $\{h_{1,2}[n],h_{2,2}[n]\}$ for $n\in\{1,4,7\}$. Since the channel values do not change over the third channel block in our model, the transmitter is able to exploit the current CSI in time slot 9, i.e., $h_{i,j}[7]=h_{i,j}[9]$. With these CSIT, transmitter 1 and transmitter 2 send a superposition of two information symbols using a proposed interference alignment technique. Using the fact that $h_{i,j}[7]=h_{i,j}[9]$, the core idea of the method is to align interference signals using current and delayed CSIT jointly so that each receiver sees the aligned interference shape that it previously obtained. For example, receiver 1 overheard unintended symbols $a_2$ and $b_2$ in a form of $L_1[4](a_2,b_2)$ in time slot 4. Similarly, receiver 2 acquired a linear combination of undesired symbols $a_1$ and $b_1$ in time slot 1, i.e., $L_2[1](a_1,b_1)$. Thus, in time slot 9, two transmitters distributively multicast four information symbols such that receiver 1 and receiver 2 observe the same interference equations $L_1[4](a_2,b_2)$ and $L_2[1](a_1,b_1)$, respectively, while providing a linearly independent equation that contains desired symbols. To satisfy this objective, transmitter 1 and transmitter 2 construct the transmit signals as
\begin{align}
x_1[9]=\frac{h_{2,1}[1]}{h_{2,1}[7]}a_1+\frac{h_{1,1}[4]}{h_{1,1}[7]}a_2, \\
x_2[9]=\frac{h_{2,2}[1]}{h_{2,2}[7]}b_1+\frac{h_{1,2}[4]}{h_{1,2}[7]}b_2.
\end{align}
Since $h_{i,j}[7]=h_{i,j}[9]$, the received signals in time slot 9 are:
\begin{align}
{ y}_{1}[9] &={h}_{1,1}[9]\left(\frac{h_{2,1}[1]}{h_{2,1}[9]}a_1+\frac{h_{1,1}[4]}{h_{1,1}[9]}a_2\right)\nonumber \\
&+  {h}_{1,2}[9]\left(\frac{h_{2,2}[4]}{h_{2,2}[9]}b_1+\frac{h_{1,2}[4]}{h_{1,2}[9]}b_2\right)\nonumber \\
&=L_1[9](a_1,b_1)+L_1[4](a_2,b_2), \\
{ y}_{2}[9] &={h}_{2,1}[9]\left(\frac{h_{2,1}[1]}{h_{2,1}[9]}a_1+\frac{h_{1,1}[4]}{h_{1,1}[9]}a_2\right)\nonumber \\
&+  {h}_{ 2,2}[9]\left(\frac{h_{2,2}[1]}{h_{2,2}[9]}b_1+\frac{h_{1,2}[4]}{h_{1,2}[9]}b_2\right)\nonumber \\
&=L_2[1](a_1,b_1)+L_2[9](a_2,b_2).
\end{align}
We explain the decoding method to recover intended symbols by focusing on receiver 1. Receiver 1 obtains a new linear combination containing desired symbols $L_1[9](a_1,b_1)$ by performing the interference cancellation, i.e., $y_1[9]-y_1[4]$.  Then, the concatenated input-output relationship is given by
\begin{align}
\left[\!\!%
\begin{array}{c}
  {y}_{1}[1]\\
 {y}_{1}[9]-y_1[4]\\
\end{array}%
\!\!\right]\!= \!\left[\!\!%
\begin{array}{cc}
 {h}_{1,1}[1] & { h}_{1,2}[1] \\
 \frac{{h}_{1,1}[9]h_{2,1}[1]}{h_{2,1}[9]} & \frac{{h}_{1,2}[9]h_{2,2}[4]}{h_{2,2}[9]} \\\end{array}%
\!\!\right]\!\!\left[\!\!%
\begin{array}{c}
 a_1 \\
 b_1 \\
\end{array}%
\!\!\right]. \label{eq:example_decoding}
\end{align}
Since the channel values were selected from a continuous distribution per each block, receiver 1 is able to recover $a_1$ and $b_1$  almost surely by applying a ZF decoder. By symmetry, receiver 2 operates in a similar fashion, which implies that a total $\frac{4}{3}$ of sum-DoF is achievable. 

\begin{remark} [Interpretation of the Proposed Transmission Method] Now we reinterpret the proposed transmission method from the perspective of higher-order message transmission techniques in \cite{Maddah-Ali:12}, which is helpful for understanding the key principle of the proposed transmission method. In the phase when the transmitters do not have knowledge of CSIT due to feedback delay, they send a information symbol per time slot, which can be interpreted as the first-order message. In the second phase when both current and delayed CSIT is available, the two transmitters distributively construct a second-order message based on their local CSIT. Unlike the two-user vector broadcast channel in \cite{Maddah-Ali:12}, the second-order message $L_2[1](a_1,b_1)+L_1[4](a_2,b_2)$ cannot be generated at neither transmitter 1 nor transmitter 2 due to their distributed nature.. To overcome this, our transmission method allows for the two transmitters to create the second-order messages distributively with the help of current CSIT knowledge in part. 
\end{remark}

\begin{remark} [Impact of Local and Delayed CSIT] This example elucidates that it is possible to achieve the optimal sum-DoF of the two-user X-channel with local and delayed CSIT if feedback delay is less than a certain fraction of channel coherence time. This implies that instantaneous and global CSIT over the three channel uses is not necessarily required to obtain the optimal sum-DoF in this channel.
\end{remark}


\begin{remark}[Reduced CSI Feedback Amount]
The other advantage of the proposed transmission method is that it possibly reduces the amount of CSI feedback by sending back precoding information instead of channel value itself. Specifically, only required information at transmitters in time slot 9 are effective channel values for precoding, i.e., $\left\{\frac{h_{2,1}[1]}{h_{2,1}[7]}, \frac{h_{1,1}[4]}{h_{1,1}[7]}\right\}$ for transmitter 1 and $\left\{ \frac{h_{2,2}[1]}{h_{2,2}[7]}, \frac{h_{1,2}[4]}{h_{1,2}[7]} \right\}$ for transmitter 2. Thus, by sending back those 4 complex values to the transmitters, it is possible to achieve the optimal sum-DoF in this network. Meanwhile, a conventional interference alignment in \cite{Cadambe_Jafar:08} requires to send back $4\times 3=12$ complex values to have global and instantaneous CSIT through a feedback link. Thus, the proposed method allows to decrease the amount of CSI feedback significantly.
\end{remark}
 

\begin{remark} [Connection with Alternating CSIT Framework in \cite{Tandon_alt:13}] The proposed transmission method can be explained through the lens of the alternating CSIT framework in \cite{Tandon_alt:13}. Since the distributed STIA requires to know the delayed and perfect CSIT for two-thirds and one-thirds of the time, respectively, it achieves the sum-DoF of $\frac{4}{3}$ with the alternating CSIT configuration of (DD, DD, PP). This connection is interesting because the proposed method is applicable in the multi-carrier system where delayed and perfect CSIT are available for two-thirds and one-thirds of the entire subcarriers.    
\end{remark}

\section{ Achievable Trade-Offs of the X-Channel }
In this section, we characterize an achievable sum-DoF region as a function of the normalized feedback delay for the $K\times 2$ X-channel. The main result is established in the following theorem. 
 
\vspace{2mm}
\begin{theorem} \label{Theorem1}
For the $K\times 2$ X-channel with local CSIT, an achievable trade-off region between sum-DoF and CSI feedback delay is given by
\begin{align}
d^{\rm X_{\rm L}}_{\Sigma}(K,2;\lambda) = \left\{
\begin{array}{l l}
 \frac{2K}{K+1}, & \quad \textrm{for} \quad 0\leq \lambda\leq \frac{2}{K+1}, \\
  -\lambda+2, & \quad \textrm{for} \quad \frac{2}{K+1}< \lambda\leq 1,\\
  1, & \quad \textrm{for} \quad \lambda\geq 1.\\
\end{array} \right.\\ \nonumber \label{eq:Th1}
\end{align}

\end{theorem}
\vspace{-10mm}

\subsection{Proof of Theorem 1}
 We prove Theorem 1 by focusing on the point of $d^{\rm X_{\rm L}}_{\Sigma}(K,2;\frac{2}{K+1})=\frac{2K}{K+1}$ because the other points connecting two points $d^{\rm X_{\rm L}}_{\Sigma}(K,2;\frac{2}{K+1})=\frac{2K}{K+1}$ and $d^{\rm X_{\rm L}}_{\Sigma}(K,2;1)=1$ is simply achievable by time sharing between the proposed transmission method and a TDMA method. To show the achievability of the point $d^{\rm X_{\rm L}}_{\Sigma}(K,2;\frac{2}{K+1})=\frac{2K}{K+1}$, we demonstrate that the transmitters can send 2$K$ independent symbols over $K+1$ channel uses.

We explain a method selecting $K+1$ channel uses while satisfying the CSI delay condition of $\lambda=\frac{2}{K+1}$. 
Our scheme operates over $n+K$ channel blocks and each of them consists of $K+1$ time slots so that the channel coherence time becomes $K+1$ time slots, $T_{\rm c}=K+1$. Furthermore, we define $\mathcal{S}_t=\{1,2,\ldots,(K+1)n+K(K+1)\}$ as a set of time slots for transmission with the cardinality $|\mathcal{S}_t|=(K+1)n+K(K+1)$. If the feedback delay is two time slots, $T_{\rm fb}=2$, then, the normalized feedback delay becomes $\lambda\!=\!\frac{2}{K+1}$. With this setup, we divide $\mathcal{S}_t$ into two subsets, $\mathcal{S}_c$ with $|\mathcal{S}_c|=(K-1)n+(K-1)K$ and $\mathcal{S}_d$ with $|\mathcal{S}_d|=2n+2K$ where $\mathcal{S}_c\cap \mathcal{S}_d = \phi$. Here, $\mathcal{S}_{c}$ is the set of time slots when the transmitter is able to access current CSI. Meanwhile, $\mathcal{S}_d$ represents the set of time slots when the transmitter has delayed CSI only. With $\mathcal{S}_{c}$ and $\mathcal{S}_{d}$, let us define a time slot set with $K+1$ elements for applying the proposed algorithm, i.e., $I_{\ell}=\{t_{\ell,1},t_{\ell,2},\ldots,t_{\ell,K},t_{\ell,K+1}\}$ where $\{t_{\ell,1},t_{\ell,2}\} \in \mathcal{S}_{d}$, $t_{\ell,j}\in \mathcal{S}_{c}$ for $j\in\{3,\ldots,K+1\}$, and $t_i$ and $t_j$ belong to different channel coherence time block if $i\neq j$. For example, as illustrated in Fig. \ref{fig:3}, when $K=2$ and $n=3$, there exists total $3n+6=15$ time resources and three index sets $I_1=\{1,4,9\}$, $I_2=\{2,5,12\}$, and $I_3=\{7,10,15\}$ are defined for the proposed transmission. For each time slot set $I_{\ell}=\{t_{\ell,1},t_{\ell,2},t_{\ell,3},\ldots,t_{\ell,K+1}\}$, $\ell\in\{1,2,\ldots,n\}$, we will show that $\frac{2K}{K+1}$ of sum-DoF is achievable. For simplicity, we omit the index $\ell$, i.e., $I_{\ell}=\{t_{1},t_{2},t_{3},\ldots,t_{K\!+\!1}\}$ in the remaining part of the paper.

%

\subsubsection{Phase One} The first phase uses two time
slots, i.e., $\mathcal{T}_1=\{t_1,t_2\}$. Since all transmitters have no channel knowledge in this
phase, each transmitter sends an information symbol without a special precoding technique. As a result, each receiver obtains one linear equation containing the $K$ desired symbols, while overhearing one equation consisting $K$ interfering information symbols. Specifically, in time slot $t_1$, transmitter $k\in\{1,2,\ldots,K\}$ sends information symbol $x_k[t_1]=s_{1,k}$ for receiver $1$. Alternatively, in time slot $t_2$, transmitter $k$ sends information symbol $x_{k}[t_2]=s_{2,k}$ for receiver $2$. If we ignore the noise, which does not affect
the DoF calculation, then the received signals are:
\begin{align}
{ y}_{1}[t_1]&= \sum_{k=1}^Kh_{1,k}[t_1]s_{1,k}=L_{1,1}[t_1],  \nonumber \\ 
{ y}_{2}[t_1]&= \sum_{k=1}^Kh_{2,k}[t_1]s_{1,k}=L_{1,2}[t_1],  \\
{ y}_{1}[t_2]&= \sum_{k=1}^Kh_{1,k}[t_2]s_{2,k}=L_{2,1}[t_2],  \nonumber \\ 
{ y}_{2}[t_2]&= \sum_{k=1}^Kh_{2,k}[t_2]s_{2,k}=L_{2,2}[t_2].
\end{align} \vspace{-0.1cm}

\subsubsection{Phase Two}

The second phase spans $K-1$ time
slots, i.e., $\mathcal{T}_2=\{t_3,t_4,\ldots, t_{K\!+\!1}\}$. In this phase, recall that the
transmitter $k$ is able to use both instantaneous CSI $\{h_{1,k}[n],h_{2,k}[n]\}$ for $n\in \mathcal{T}_2$ and delayed CSI $\{h_{1,k}[n],h_{2,k}[n]\}$ for $n\in \mathcal{T}_1$. With both outdated and current CSI knowledge, in time slot $n\in\mathcal{T}_{2}$, transmitter $k$ repeatedly sends the superposition of two independent symbols $s_{1,k}$ and $s_{2,k}$ using the precoding coefficients $v_{1,k}[n]$ and $v_{2,k}[n]$, which change over time index $n$. Then, the transmitted signal at time slot $n$ is given by
\begin{align}
{ x}_k[n]={v}_{1,k}[n]{ s}_{1,k}+{v}_{2,k}[n]{ s}_{2,k}, \label{eq:STIAphase2send}
\end{align}
where $n\in\mathcal{T}_{2}$. When the transmitters send the signal in (\ref{eq:STIAphase2send}) in time slot $n$, receiver $\ell\in\{1,2\}$ obtains
\begin{align}
{ y}_{\ell}[n]&=\sum_{k=1}^Kh_{\ell,k}[n]x_k[n] \nonumber \\
&= \underbrace{\sum_{k=1}^Kh_{\ell,k}[n]v_{1,k}[n]s_{1,k}}_{{ L}_{\ell,1}[n]}\!+\!\underbrace{\sum_{k=1}^Kh_{\ell,k}[n]v_{2,k}[n]s_{2,k}}_{{ L}_{\ell,2}[n]}.
\end{align} 
The key principle for designing the precoding coefficients ${v}_{1,k}[n]$ and ${v}_{2,k}[n]$ is to ensure that each receiver obtains the additional $K-1$ linearly independent equations that contain the desired symbols while repeatedly providing the same linear equation that contains the interfering symbols, i.e., aligning interference over different channel uses. For instance, receiver 1 has acquired one desired equation $L_{1,1}[t_1]$ and overheard one interfering equation $L_{1,2}[t_2]$ through phase one.  Therefore, if receiver $1$ repeatedly receives the same linear combinations of the interfering symbols in time slot $n$, i.e., $L_{1,2}[n]=L_{1,2}[t_2]$ for $n\in \mathcal{T}_2$, then it is possible to eliminate the effect of the sum of interfering symbols from $y_1[n]$ by subtracting the previously overheard signal $L_{1,2}[t_2]$
as side-information. To satisfy this,  we choose the precoding coefficients ${v}_{1,k}[n]$ and ${v}_{2,k}[n]$ as
\begin{align}
h_{2,k}[n]{v}_{1,k}[n]= h_{2,k}[t_1], \nonumber \\
h_{1,k}[n]{v}_{2,k}[n]= h_{1,k}[t_2], \label{STIA_cond}
\end{align}
where $k\in\{1,2,\ldots,K\}$, $n\in\{t_3,t_4, \ldots, t_{K\!+\!1}\}$. This inter-user interference
alignment condition enables that information symbols for receiver $1$ (2) have the same interference shape with the linear combination of the interference signal overheard by receiver 2 (1) in time slot $t_1$ $(t_2)$ in the first phase. Since channel $h_{\ell,k}[n]$ was selected from a continuous distribution, we construct the precoding coefficients satisfying the conditions in (\ref{STIA_cond}) as 
\begin{align}
{v}_{1,k}[n]= \frac{h_{2,k}[t_1]}{h_{2,k}[n]} \quad  {\rm and} \quad
{v}_{2,k}[n]= \frac{h_{1,k}[t_2]}{h_{1,k}[n]}.
\label{eq:STIA_BF}
\end{align} 

Next, we consider decodability. To make the exposition concrete, we focus on the decoding process at receiver 1. The decoding method involves two steps: 1) aligned interference cancellation and 2) a zero-forcing method for the extraction of the desired symbols. 

Using the fact that receiver 1 received the same linear combination of interfering symbols that it obtained during the phase two, i.e., $L_{1,2}[n]=L_{1,2}[t_2]$ for $n\in\mathcal{T}_{2}$, receiver 1 performs the aligned interference cancellation to extract the equations that contain the desired symbols as
\begin{align}
y_1[n]-y_1[t_2]&=L_{1,1}[n]+L_{1,2}[n]-L_{1,2}[t_2], \nonumber \\
&=L_{1,1}[n], \nonumber \\
&=\sum_{k=1}^K \underbrace{h_{1,k}[n]v_{1,k}[n]}_{\tilde{h}_{1,k}[n]}s_{1,k}.
\end{align}  
After the aligned interference cancellation, we obtain the concatenated system of equations 
for receiver 1 as 
\begin{align}
\left[\!\!\!%
\begin{array}{c}
  {y}_1[t_1] \\ \hline
  y_1[t_3]-y_1[t_2] \\
  \vdots \\
 y_1[t_{K\!+\!1}]-y_1[t_2]\\
\end{array}\!\!\!
\right]
&\!\!\!=\!\!\! \underbrace{ \left[\!\!\
\begin{array}{ccc}
   h_{1,1}[t_1]  &  \cdots \!\!\!\! &\!\!\!\! h_{1,K}[t_1] \\ \hline
   \tilde{h}_{1,1}[t_3]  & \cdots  \!\!\!\! &\!\!\!\!  \tilde{h}_{1,K}[t_3] \\ 
  \vdots         & \ddots  \!\!\!\! &\!\!\!\!  \vdots\\
   \tilde{h}_{1,1}\![t_{K\!+\!1}\!]   & \cdots  &\!\!\!\!  \tilde{h}_{1,K}[t_{K\!+\!1}\!] \\ 
\end{array}\!\!\!%
\right]}_{{\bf \hat H}_1}\!\!\!\!
\left[\!\!\!%
\begin{array}{c}
  {s}_{1,1} \\ 
  s_{1,2} \\
  \vdots \\
 s_{1,K}\\
\end{array}\!\!\!
\right]. \nonumber
\end{align}
Since we pick the precoding coefficients $v_{1,k}[n]$ independent of the channel value $h_{1,k}[n]$, the effective channel matrix
for ${\bf {\hat{H}}}_{1}$ has a full rank of
$K$ almost surely, thereby in the high \textrm{SNR} regime, receiver 1 is able to  decode the desired $K$ symbols 
$\{s_{1,1},\ldots,s_{1,K}\}$ by applying a ZF decoder. By symmetry, it is possible to obtain $K$ data symbols over
$K+1$ time slots for receiver 2, provided that the feedback delay is less than the $\frac{2}{K+1}$ fraction of the channel coherence time. As a result, a total $2K$ sum-DoF is achievable over $|I_{\ell}|=K+1$ channel uses.

Recall that a total number of channel uses was $|\mathcal{S}_t|=(K+1)n+K(K+1)$ and 
we have shown that $2Kn$ sum-DoF are achievable over $n(K+1)$ time slots, i.e., $|I_1\cup I_2 \cup \cdots \cup I_n|=(K+1)n$. For the residual time resources $|\mathcal{S}_t-\{I_1\cup I_2 \cup \cdots \cup I_n\}| =K(K+1)$, we simply apply a TDMA transmission method achieving $K(K+1)$ sum-DoF. Then, as $n$ goes to infinity, the asymptotically achievable sum DoF is given as:
\begin{align}
\lim_{n\rightarrow \infty}d^{X_{\rm L}}_{\Sigma}\left(K,2;\frac{2}{K+1}\right)&=\!\frac{ 2Kn +K(K+1) }{(K+1)n+K(K+1)}, \nonumber\\
&=\frac{2K}{K+1},
\end{align}
which completes the proof.\qed
\vspace{0.2cm}

Now, we make several remarks on the implication of our results.

\begin{remark}[An Extension to the MIMO X-channel] With the proposed achievability method used for proving Theorem 1, one can easily prove that the optimal sum-DoF of $\frac{2KM}{K+1}$ is achievable for the $K\times 2$ MIMO X-channel with $M$ antennas at each node almost surely when the transmitters have local CSI and the normalized feedback delay is less than $\frac{2}{K+1}$.\vspace{0.1cm}
\end{remark}

\begin{remark}[A Lower Bound of $K\times N$ X-channel] From the fact that  any achievable sum-DoF in the $K\times 2$ X-channel is also achievable in the $K\times N$ X-channel for $N\geq 2$, we are able to establish a lower bound of the sum-DoF region for $K\times N$ X-channel as $d^{X_{\rm L}}_{\Sigma}(K,N;\lambda) \geq d^{X_{\rm L}}_{\Sigma}(K,2;\lambda)$. The lower bound does not scale with neither $K$ or $N$, the number of transmitters or receivers. Nevertheless, the lower bound is strictly better than the best known lower bound for the case with delayed CSI alone [9] for all values of $K$. \vspace{0.1cm}
\end{remark}

\begin{remark}[CSI Feedback Delay] The proposed method achieves the optimal sum-DoF for $K \times 2$ X-channel with local CSIT as long as the normalized feedback delay is less than $\frac{2}{K+1}$. This does not necessarily imply that the maximum allowable normalized feedback delay achieving the optimal sum-DoF is $\frac{2}{K+1}$. In other words, we do not establish any optimality claim on our achievable sum-DoF region with respective to the normalized feedback delay. The problem of characterizing the maximum allowable feedback delay remains an interesting open problem. 
\end{remark}

\begin{remark}[An Extension to the Two-Cell Interfering Multiple Access Channel] Let us consider an analogous network in which $K$ users (transmitters) per cell intend to communicate to their respective base station (receiver) while interfering with each other. In particular, when the number of cells is two, then we refer this network to a two-cell interfering multiple access channel. In this network, one can easily apply the proposed method to show that the sum-DoF of $\frac{2K}{K+1}$ is achievable with local CSIT if the normalized feedback delay is less than $\frac{2}{K+1}$.\vspace{0.1cm}
\end{remark}

\subsection{Comparison of Achievable Trade-Offs }
To shed further light on the significance of the trade-off region derived in Theorem 1, it is instructive to compare it with the other regions achieved by different methods when $K=2$. For the two-user X-channel, by leveraging the transmission method proposed in \cite{GMK_X:11}, it is possible to establish a trade-off region for the $2\times 2$ X-channel with global and delayed CSIT, which is stated in the following corollary. 

\begin{figure}
\centering
\includegraphics[width=3.3in]{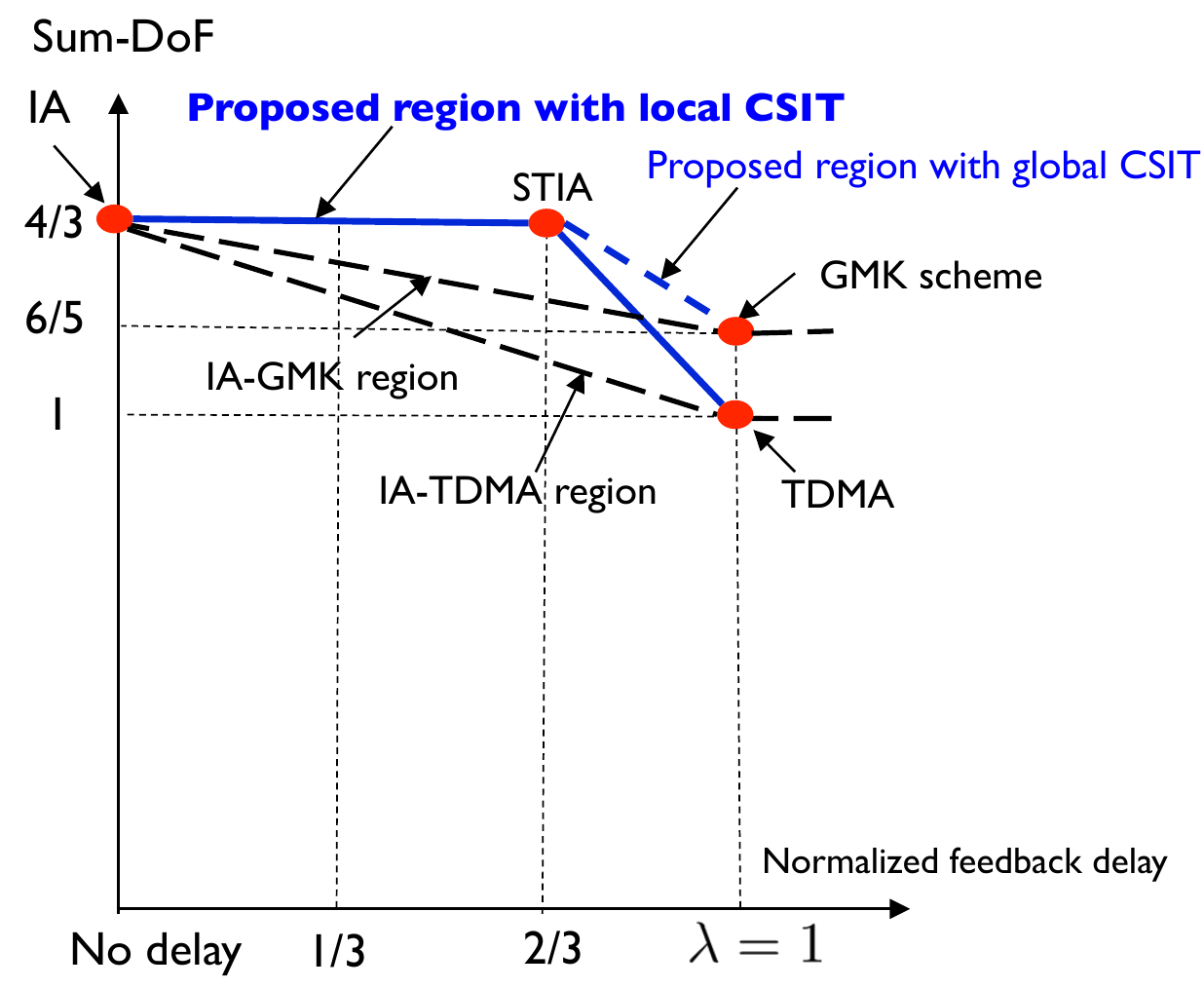}
\caption{Illustration of trade-offs for the 2$\times$2 X-channel.} \label{fig:4}
\end{figure}
%
%
%

\begin{corollary} \label{Corollary1}
A CSI feedback delay-DoF gain trade-off region for the two-user X-channel with global and delayed CSIT is given by
\begin{align}
d^{X_{\rm G}}_{\Sigma}(2,2;\lambda) = \left\{
\begin{array}{l l}
  \frac{4}{3}, & \quad \textrm{for} \quad 0\leq \lambda\leq \frac{2}{3}, \\
  -\frac{2}{5}\lambda+\frac{8}{5}, & \quad \textrm{for} \quad \frac{1}{3}< \lambda\leq 1,\\
  \frac{6}{5}, & \quad \textrm{for} \quad \lambda\geq 1.\\
\end{array} \right.\\ \nonumber \label{eq:co1}
\end{align}
\end{corollary}

\textit{Proof:} Achievability of the trade-off region is direct from Theorem 1 and the AGK method proposed in \cite{GMK_X:11}. For the case of $\lambda \leq \frac{2}{3}$, from Theorem 1, it was shown that $d^{\rm X_G}_{\Sigma}\left(2,2;\lambda\right)=\frac{4}{3}$ of sum-DoF are achievable. Alternatively, for the completely-delayed regime, $\lambda \geq 1$,  the transmission method in \cite{GMK_X:11} exploiting global CSIT allows to attain $d^{\rm X_G}_{\Sigma}\left(2,2;1\right)=\frac{6}{5}$ of sum-DoF. As a result, it is possible to achieve any points in the line connecting two points between $d^{\rm X_G}_{\Sigma}\left(2,2;\frac{2}{3}\right)$, and $d^{\rm X_G}_{\Sigma}\left(2,2;1\right)$ through a time-sharing technique. \qed  \vspace{0.2cm}

Using a time sharing technique between IA and TDMA, the region of $d_{\textrm{IA-TDMA}}^{X_{\rm G}}(2,2;\lambda)=-\frac{\lambda}{3}+\frac{4}{3}$ is achievable with global CSIT for $0\leq \lambda \leq 1$. Analogously, if we apply the time sharing method between IA and GMK method, then the region of $d_{\textrm{IA-GMK}}^{X_{\rm G}}(2,2;\lambda)=-\frac{2}{15}\lambda+\frac{4}{3}$ is achieved with global CSIT for $0\leq \lambda \leq 1$. Therefore, as illustrated in Fig. \ref{fig:4}, the proposed method allows to attain a higher trade-off region between CSI feedback delay and sum-DoF than those obtained by the other methods when the CSI feedback is not too delayed. For example, when $\lambda=\frac{2}{3}$, the proposed method achieves the sum-DoF of $\frac{4}{3}$, which yields the $\frac{2}{9}$ sum-DoF gain over IA-TDMA and $\frac{4}{45}$ sum-DoF gain over IA-GMK even with local CSIT. Another remarkable point is that CSIT sharing between transmitters does not contribute to improve the sum-DoF if the feedback delay is less than two-thirds of the channel coherence time. Whereas, global CSIT knowledge is useful to increase the DoF performance as the normalized feedback delay increases beyond $\lambda >\frac{2}{3}$.
 \vspace{0.1cm}

\section{ An Achievable Trade-Offs \\of the Three-User Interference Channel} 
In this section we characterize the trade-off region between the sum-DoF and CSI feedback delay for the three-user interference channel with local and delayed CSIT. In this channel, designing interference alignment algorithm with local and delayed CSIT is more challenging than the case of the $K\times 2$ X-channel. This difficulty comes from that it may be impossible for simultaneously aligning interference at more than two receivers in a distributed manner. Interestingly, even in this setting, we show that local and delayed CSIT still provides a gain in DoF beyond that obtained by TDMA. The following is the main result of this section.

\begin{figure}
\centering
\includegraphics[width=3.5in]{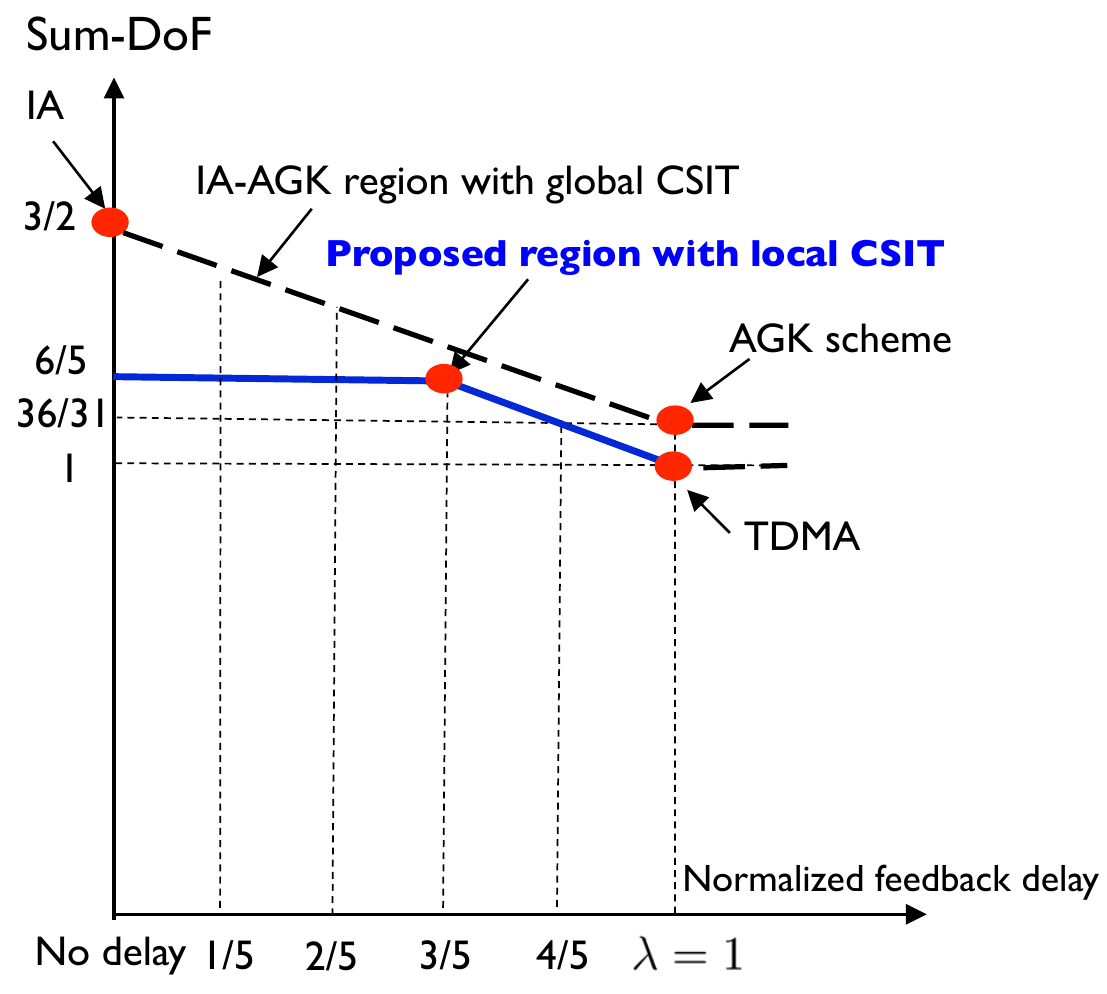}
\caption{Illustration of trade-offs for the three-user interference channel.} \label{fig:5}
\end{figure}

\begin{theorem} \label{Theorem2}
For the three-user interference channel with distributed and delayed CSIT, the trade-off region between the sum-DoF and the feedback delay is  
\begin{align}
d^{\rm IC_{\rm L}}_{\Sigma}(3,3;\lambda) = \left\{
\begin{array}{l l}
  \frac{6}{5}, & \quad \textrm{for} \quad 0\leq \lambda\leq \frac{3}{5}, \\
  -\frac{1}{2}\lambda+\frac{3}{2}, & \quad \textrm{for} \quad \frac{3}{5}< \lambda\leq 1,\\
  1, & \quad \textrm{for} \quad \lambda\geq 1.\\
\end{array} \right.\\ \nonumber \label{eq:Th2}
\end{align}
\end{theorem}
\vspace{-0.1cm}
\subsection{Proof of Theorem 2}
We focus on proving the point of $d_{\Sigma}^{\rm IC_L}(3,3;\frac{3}{5})=\frac{6}{5}$ because other points connecting $d_{\Sigma}^{\rm IC_L}(3,3;\frac{3}{5})=\frac{6}{5}$ and $d_{\Sigma}^{\rm IC_L}(3,3;1)=1$ are simply attained by time sharing between the proposed method and TDMA transmission. To show the achievability of $d_{\Sigma}^{\rm IC_L}(3,3;\frac{3}{5})=\frac{6}{5}$, we demonstrate that a total of six independent symbols can be reliably delivered over five time slots. Without loss of generality, we assume that transmitter $k\in\{1,2,3\}$ is able to access local and delayed CSIT over three time slots $\{h_{1,k}[n],h_{2,k}[n],h_{3,k}[n]\}$ for $n\in\{1,2,3\}$, while having local and current CSIT over two time slots, $\{h_{1,k}[n],h_{2,k}[n],h_{3,k}[n]\}$ for $n\in\{4,5\}$ under the premise of $\lambda=\frac{3}{5}$. 

\subsubsection{Phase One}
Phase one spans three time slots. In this phase, a scheduled transmission is applied, which requires no CSIT. Specifically, in time slot 1, transmitter 1 and transmitter 2 send information symbol $a_{1}$ and $b_{1}$. Then, the received signals are 
\begin{align}
{ y}_{1}[1]&= h_{1,1}[1]a_1+h_{1,2}[1]b_1=L_{1}[1](a_1,b_1),  \nonumber \\ 
{ y}_{2}[1]&= h_{2,1}[1]a_1+h_{2,2}[1]b_1=L_{2}[1](a_1,b_1),  \nonumber \\ 
{ y}_{3}[1]&= h_{3,1}[1]a_1+h_{3,2}[1]b_1=L_{3}[1](a_1,b_1).
\end{align} 
In time slot 2, transmitter 1 and transmitter 3 send symbol $a_2$ and $c_1$. Then, the received signals are 
\begin{align}
{ y}_{1}[2]&= h_{1,1}[2]a_2+h_{1,3}[2]c_1=L_{1}[2](a_2,c_1),  \nonumber \\ 
{ y}_{2}[2]&= h_{2,1}[2]a_2+h_{2,3}[2]c_1=L_{2}[2](a_2,c_1),  \nonumber \\ 
{ y}_{3}[2]&= h_{3,1}[2]a_2+h_{3,3}[2]c_1=L_{3}[2](a_2,c_1).
\end{align} 
In time slot 3, transmitter 2 and transmitter 3 send information symbol $b_2$ and $c_2$. Then, the received signals are 
\begin{align}
{ y}_{1}[3]&= h_{1,2}[3]b_2+h_{1,3}[3]c_2=L_{1}[3](b_2,c_2),  \nonumber \\ 
{ y}_{2}[3]&= h_{2,2}[3]b_2+h_{2,3}[3]c_2=L_{2}[3](b_2,c_2),  \nonumber \\ 
{ y}_{3}[3]&= h_{3,2}[3]b_2+h_{3,3}[3]c_2=L_{3}[3](b_2,c_2).
\end{align} 

\subsubsection{Phase Two}
Phase two uses two time slots. Recall that, in the second phase, transmitters exploit both current and outdated CSIT. In time slot 4, transmitter 1 sends a superposition of $a_1$ and $a_2$ with the precoding coefficients $v_{a,1}[4]$ and $v_{a,2}[4]$; transmitter 2 and transmitter 3 send $b_1$ and $c_1$ with the precoding coefficients $v_{b,1}[4]$ and $v_{c,1}[4]$. The construction method of the precoding coefficients is to provide the same interference shape to receivers what they previously obtained during the phase one in a distributed manner. Specifically, the precoding coefficients are chosen as $v_{a,1}[4]=\frac{h_{3,1}[1]}{h_{3,1}[4]}$, $v_{a,2}[4]=\frac{h_{2,1}[2]}{h_{2,1}[4]}$, $v_{b,1}[4]=\frac{h_{3,2}[1]}{h_{3,2}[4]}$, and $v_{c,1}[4]=\frac{h_{2,3}[2]}{h_{2,3}[4]}$. This allows for receiver 2 and 3 to obtain the aligned interference shape that they acquired in time slot 1 and time slot 2, respectively. Then, the received signals are
\begin{align}
{ y}_{1}[4]&= h_{1,1}[4](v_{a,1}[4]a_1\!+\!v_{a,2}[4]a_2) \nonumber \\ &+h_{1,2}[4]v_{b,1}[4]b_1  +h_{1,3}[4]v_{c,1}[4]c_1,\nonumber \\ &=L_{1}[4](a_1,b_1)+L_{1}[4](a_2,c_1), \nonumber \\ 
{ y}_{2}[4]&= h_{2,1}[4](v_{a,1}[4]a_1\!+\!v_{a,2}[4]a_2) \nonumber \\ &+h_{2,2}[4]v_{b,1}[4]b_1  +h_{2,3}[4]v_{c,1}[4]c_1,\nonumber \\ &=L_{2}[4](a_1,b_1)+L_{2}[2](a_2,c_1),  \nonumber \\ 
{ y}_{3}[4]&= h_{3,1}[4](v_{a,1}[4]a_1\!+\!v_{a,2}[4]a_2) \nonumber \\ &+h_{3,2}[4]v_{b,1}[4]b_1  +h_{3,3}[4]v_{c,1}[4]c_1,\nonumber \\ &=L_{3}[1](a_1,b_1)+L_{3}[4](a_2,c_1).
\end{align} 
In time slot 5, transmitter 2 sends a linear combination of $b_1$ and $b_2$ using the precoding coefficients $v_{b,1}[5]$ and $v_{b,2}[5]$; transmitter 1 and transmitter 3 send information symbol $a_1$ and $c_2$ by applying the precoding $v_{a,1}[5]$ and $v_{c,2}[5]$, respectively. In particular, the precoding coefficients are selected as $v_{b,1}[5]=\frac{h_{3,2}[1]}{h_{3,2}[5]}$, $v_{b,2}[5]=\frac{h_{1,2}[3]}{h_{1,2}[5]}$, $v_{a,1}[5]=\frac{h_{3,1}[1]}{h_{3,1}[5]}$, and $v_{c,2}[5]=\frac{h_{1,3}[3]}{h_{1,3}[5]}$ so that receiver 1 receives the aligned interference shape with what it obtained in time slot 3. Then, the received signals are
\begin{align}
{ y}_{1}[5]&= h_{1,1}[5]v_{a,1}[5]a_1 +h_{1,2}[5]v_{b,1}[5]b_1\nonumber \\ &+h_{1,2}[5]v_{b,2}[5]b_2  +h_{1,3}[5]v_{c,2}[5]c_2,\nonumber \\ &=L_{1}[5](a_1,b_1)+L_{1}[3](b_2,c_2), \nonumber \\ 
{ y}_{2}[5]&= h_{2,1}[5]v_{a,1}[5]a_1 +h_{2,2}[5]v_{b,2}[5]b_1\nonumber \\ &+h_{2,2}[5]v_{b,2}[5]b_2 +h_{2,3}[5]v_{c,2}[5]c_2,\nonumber \\ &=L_{2}[5](a_1,b_1)+L_{2}[5](b_2,c_2),  \nonumber \\ 
{ y}_{3}[5]&= h_{3,1}[5]v_{a,1}[5]a_1 +h_{3,2}[5]v_{b,2}[5]b_1\nonumber \\ &+h_{3,2}[5]v_{b,2}[5]b_2 +h_{3,3}[5]v_{c,2}[5]c_2,\nonumber \\ &=L_{3}[1](a_1,b_1)+L_{3}[5](b_2,c_2).
\end{align}

Now, we explain how each receiver decodes its two desired symbols through a successive interference cancellation technique.
\begin{itemize}
\item Receiver 1 first obtains a linear combination of $a_1$ and $b_1$ by subtracting $y_1[3]$ from $y_1[5]$, i.e., $y_1[5]-y_1[3]=L_1[5](a_1,b_1)$. Then, combining $y_1[1]=L_1[1](a_1,b_1)$ and $L_1[5](a_1,b_1)$, receiver 1 decodes both $a_1$ and $b_1$. Using decoded symbols $a_1$ and $b_1$, receiver 1 obtains a linear equation $L_1[4](a_2,c_1)$ from $y_1[4]$. Lastly, concatenating  $y_1[2]=L_1[2](a_2,c_1)$ and $L_1[4](a_2,c_1)$, receiver 1 resolves $a_2$ and $c_1$. Therefore, receiver 1 decodes two desired symbols $a_1$ and $a_2$.

\item Receiver 2 cancels the effect of interference symbols $a_2$ and $c_1$ in $y_2[4]$ using side information obtained in time slot 2, i.e., $y_2[2]$, thereby receiver 2 extracts a linear combination of $a_1$ and $b_1$ as
\begin{align}
y_2[4]-y_2[2]&=L_2[4](a_1,b_1)+L_2[2](a_2,c_1)\nonumber \\ &-L_2[2](a_2,c_1) 
=L_2[4](a_1,b_1).
\end{align}
Combining the received equation in time slot 1 $L_2[1](a_1,b_1)$ with $L_2[4](a_1,b_1)$, receiver 2 decodes both $a_1$ and $b_1$. Using decoded symbols $a_1$ and $b_1$ and the effective channel values, it creates $L_2[5](a_1,b_1)$. Subtracting $L_2[5](a_1,b_1)$ from $y_2[5]$, it is possible to obtain a new linear equation that contains information symbols $b_2$ and $c_2$ only, i.e., $L_2[5](b_2,c_2)$. Finally, using both $L_2[5](b_2,c_2)$ and $L_2[3](b_2,c_2)$, receiver 2 decodes $b_2$ and $c_2$.

\item Receiver 3 removes the effect of interference symbols $a_1$ and $b_1$ in $y_3[5]$ using side information acquired in time slot 1, i.e., $y_3[1]$, thereby receiver 2 extracts a linear combination of $b_2$ and $c_2$ as
\begin{align}
y_3[5]-y_3[1]&=L_3[5](a_1,b_1)+L_3[5](b_2,c_2)\nonumber \\ &-L_3[2](a_1,b_1) 
=L_3[5](b_2,c_2).
\end{align}
Putting the received equations in time slot 3 and 5, $L_3[3](b_2,c_2)$ and $L_3[5](b_2,c_2)$ together, receiver 3 retrieves both $b_2$ and $c_2$. Further, receiver 2 subtracts $y_3[1]=L_3[1](a_1,b_1)$ from $y_3[4]$, which provides a new linear equation $L_3[4](a_2,c_1)$. Since receiver 3 already obtained a different  equation $L_3[2](a_2,c_1)$, it is possible to decode both $a_2$ and $c_1$. 
\end{itemize}
\vspace{0.1cm}

\begin{remark}[Sensitivity of Feedback Delay]  Unlike the $2\times 2$ X-channel, the sum-DoF loss occurs significantly due to the CSI feedback delay in the three-user interference channel. Nevertheless, the sum-DoF stated in Theorem 2, $d_{\Sigma}^{\rm IC_L}(3,3;\frac{3}{5})=\frac{6}{5}$, is strictly higher than the best known sum-DoF result $d_{\Sigma}^{\rm IC_G}(3,3;1)=\frac{36}{31}$ for the case with global and completely-delayed CSI alone, i.e., $\lambda\geq 1$ in \cite{AGK:13}.  \end{remark}

\section{Achievable Rate Analysis \\of the $2\times 2$ X-channel }
In the previous sections, we characterized the trade-off between the sum-DoF and CSI feedback delay, assuming high SNR and showed that the proposed method improves the sum-DoF in the X-channel and a three-user interference channel with moderately-delayed and local CSIT. In this section, instead of a DoF metric, we analyze the achievable rate of the proposed interference alignment at the finite SNR, particularly focusing on the $2\times 2$ X-channel with moderately-delayed CSIT. Leveraging an information theoretic outer bound in \cite{Cadambe_Jafar:09}, together with the derived achievable rate, we demonstrate that the proposed method achieves the sum-capacity of the two-user X-channel within a constant gap over the all range of SNR and $\lambda \leq \frac{2}{3}$ for a particular channel setting.

\begin{theorem} \label{Theorem3}
For the two-user X-channel with $\lambda \leq \frac{2}{3}$, the achievable sum-rate is given by  
\begin{align}
R_{\Sigma}^{X}(\lambda)= \frac{ \sum_{k=1}^2\log_2\left[ \det \left({\bf I} + P~{\bf {H}}_{k} {\bf {Z}}_k^{-1} {\bf {H}}_{k}^{*} \right)\right]}{3}, \label{eq:Th3}
\end{align}
where
\begin{align}
{\bf {H}}_{1} &=\left[\!\!\!%
\begin{array}{cc}
 {h}_{1,1}[1] & { h}_{1,2}[1] \\
 \frac{{h}_{1,1}[9]h_{2,1}[1]}{h_{2,1}[9]} &\frac{{h}_{1,2}[9]h_{2,2}[4]}{h_{2,2}[9]} \\\end{array}%
\!\!\!\!\right], \nonumber \\{\bf H}_2&=\left[\!\!%
\begin{array}{cc}
 {h}_{2,1}[4] & { h}_{2,2}[4] \\
 \frac{{h}_{2,1}[9]h_{1,1}[4]}{h_{1,1}[9]} & \frac{{h}_{2,2}[9]h_{1,2}[4]}{h_{1,2}[9]} \\\end{array}%
\!\!\!\right] \nonumber,
\end{align}
\begin{align}
{\bf Z}_1=\sigma^2 \left[%
\begin{array}{cc}
 1 & 0 \\
 0 &  1+\frac{1}{p_{1}^{\star}} \\
\end{array}%
\right], ~~{\bf Z}_2= \sigma^2 \left[%
\begin{array}{cc}
 1 & 0 \\
 0 & 1+\frac{1}{p_{2}^{\star}} \\
\end{array}%
\right],  \nonumber
\end{align} and
\begin{align}
\left[%
\begin{array}{cc}
\left|\frac{h_{2,1}[1]}{h_{2,1}[9]}\right|^2 &\left|\frac{h_{1,1}[4]}{h_{1,1}[9]}\right|^2 \\
 \left|\frac{h_{2,2}[1]}{h_{2,2}[9]}\right|^2& \left|\frac{h_{1,2}[4]}{h_{1,2}[9]}\right|^2\
 \\\end{array}%
\right]\left[\!\!%
\begin{array}{c}
 p_{1}^{\star} \\
 p_{2}^{\star} \\
\end{array}%
\!\!\right] \leq \left[\!\!%
\begin{array}{c}
 1 \\
 1\\
\end{array}%
\!\!\right]. \nonumber
\end{align}
\end{theorem}

 \proof

From the proposed transmission method explained in Section II, the received signals during time slot 1 and 4 are  
\begin{align}
{ y}_{1}[1] &={h}_{1,1}[1]a_1+  {h}_{2,2}[1]b_1+z_1[1],  \\
{ y}_{2}[1] &={h}_{2,1}[1]a_1+  {h}_{2,2}[1]b_1+z_2[1], \\
{ y}_{1}[4] &={h}_{1,1}[4]a_2+  {h}_{2,2}[4]b_2+z_1[4], \\
{ y}_{2}[4] &= {h}_{2,1}[4]a_2+  {h}_{2,2}[4]b_2+z_2[4].
\end{align}
In time slot 9, using the current and outdated CSIT, transmitter 1 and 2 send signals as\begin{align}
x_1[9]=\sqrt{p_{1,1}}\frac{h_{2,1}[1]}{h_{2,1}[9]}a_1+\sqrt{p_{2,1}}\frac{h_{1,1}[4]}{h_{1,1}[9]}a_2, \\
x_2[9]=\sqrt{p_{1,2}}\frac{h_{2,2}[1]}{h_{2,2}[9]}b_1+\sqrt{p_{2,2}}\frac{h_{1,2}[4]}{h_{1,2}[9]}b_2.
\end{align}
where $p_{1,1}$, $p_{1,2}$, $p_{2,1}$, and $ p_{2,2}$ denote the transmit power coefficients carrying information symbols, $a_1$, $a_2$, $b_1$, and $b_2$. To ensure the transmit power constraints, we need to pick $p_{i,j}$ such that
\begin{align}
&p_{1,1}\left|\frac{h_{2,1}[1]}{h_{2,1}[9]}\right|^2+p_{2,1}\left|\frac{h_{1,1}[4]}{h_{1,1}[9]}\right|^2\leq 1 \label{eq:pwcont1} \\ 
&p_{1,2}\left|\frac{h_{2,2}[1]}{h_{2,2}[9]}\right|^2+p_{2,2}\left|\frac{h_{1,2}[4]}{h_{1,2}[9]}\right|^2\leq 1. \label{eq:pwcont2}
\end{align}
Further, for the interference alignment condition, we need to impose the additional constraints such that $p_{1,1}=p_{1,2}$ and $p_{2,1}=p_{2,2}$. Therefore, the power coefficients satisfying the constraints are obtained as 
\begin{align}
\left[\!\!%
\begin{array}{c}
 p_{1}^{\star} \\
 p_{2}^{\star} \\
\end{array}%
\!\!\right] = \left[%
\begin{array}{cc}
\left|\frac{h_{2,1}[1]}{h_{2,1}[9]}\right|^2 &\left|\frac{h_{1,1}[4]}{h_{1,1}[9]}\right|^2 \\
 \left|\frac{h_{2,2}[1]}{h_{2,2}[9]}\right|^2& \left|\frac{h_{1,2}[4]}{h_{1,2}[9]}\right|^2\
 \\\end{array}%
\right]^{-1} \left[\!\!%
\begin{array}{c}
 1 \\
 1\\
\end{array}%
\!\!\right].
\end{align}
Then, the received signals in time slot 9 are:
\begin{align}
{ y}_{1}[9] &={h}_{1,1}[9]\left( \sqrt{p_{1}^{\star}}\frac{h_{2,1}[1]}{h_{2,1}[9]}a_1+\sqrt{p_{2}^{\star}}\frac{h_{1,1}[4]}{h_{1,1}[9]}a_2\right)\nonumber \\
&\!\!\!\!\!+\!\  {h}_{1,2}[9]\left(\sqrt{p_{1}^{\star}}\frac{h_{2,2}[1]}{h_{2,2}[9]}b_1+\sqrt{p_{2}^{\star}}\frac{h_{1,2}[4]}{h_{1,2}[9]}b_2\right)\!+\!\!z_1[9]\nonumber \\
&=\sqrt{p_{1}^{\star}}L_1[9](a_1,b_1)+\sqrt{p_{2}^{\star}}L_1[4](a_2,b_2)+z_1[9], \\
{ y}_{2}[9] &={h}_{2,1}[9]\left(\sqrt{p_{1}^{\star}}\frac{h_{2,1}[1]}{h_{2,1}[9]}a_1+\sqrt{p_{2}^{\star}}\frac{h_{1,1}[4]}{h_{1,1}[9]}a_2\right)\nonumber \\
&\!\!\!\!+\!  {h}_{ 2,2}[9]\left(\sqrt{p_{1}^{\star}}\frac{h_{2,2}[1]}{h_{2,2}[9]}b_1+\sqrt{p_{2}^{\star}}\frac{h_{1,2}[4]}{h_{1,2}[9]}b_2\right)\!+\!z_2[9]\nonumber \\
&=\sqrt{p_{1}^{\star}}L_2[1](a_1,b_1)+\sqrt{p_{2}^{\star}}L_2[9](a_2,b_2)+z_2[9].
\end{align}
Applying the interference cancellation, i.e., $y_1[9]-\sqrt{p_{2}^{\star}}y_1[4]$ and multiplying normalization factor $\frac{1}{\sqrt{p_2^{\star}}}$, receiver 1 has the following resultant input-output relationship:
\begin{align}
\left[%
\begin{array}{c}
  {y}_{1}[1]\\
 \frac{{y}_{1}[9]}{\sqrt{p_{2}^{\star}}}-y_1[4]\\
\end{array}%
\right]&\!=\!\! \underbrace{\left[%
\begin{array}{cc}
 {h}_{1,1}[1] & { h}_{1,2}[1] \\
\frac{{h}_{1,1}[9]h_{2,1}[1]}{h_{2,1}[9]} &\frac{{h}_{1,2}[9]h_{2,2}[4]}{h_{2,2}[9]} \\\end{array}%
\right]}_{{\bf H}_1}\!\!\left[\!\!%
\begin{array}{c}
 a_1 \\
 b_1 \\
\end{array}%
\!\!\right] \nonumber \\&+\underbrace{\left[\!\!%
\begin{array}{c}
 z_1[1] \\
  \frac{{z}_{1}[9]}{\sqrt{p_{2}^{\star}}}-z_{1}[4] \\
\end{array}%
\!\!\right]}_{{\bf z}_1}. \label{eq:sumrate1}
\end{align} 
Similarly, the resulting input-output relationship at receiver 2 is 
\begin{align}
\left[%
\begin{array}{c}
  {y}_{2}[4]\\
 \frac{{y}_{2}[9]}{\sqrt{p_{1}^{\star}}}-y_2[1]\\
\end{array}%
\right]&\!=\!\! \underbrace{\left[%
\begin{array}{cc}
 {h}_{2,1}[4] & { h}_{2,2}[4] \\
\frac{{h}_{2,1}[9]h_{1,1}[4]}{h_{1,1}[9]} & \frac{{h}_{2,2}[9]h_{1,2}[4]}{h_{1,2}[9]} \\\end{array}%
\right]}_{{\bf H}_2}\!\!\left[\!\!%
\begin{array}{c}
 a_2\\
 b_2 \\
\end{array}%
\!\!\right] \nonumber \\&+\underbrace{\left[\!\!%
\begin{array}{c}
 z_2[4] \\
\frac{{z}_{2}[9]}{\sqrt{p_{1}^{\star}}}-z_{2}[1] \\
\end{array}%
\!\!\right]}_{{\bf z}_2}. \label{eq:sumrate2}
\end{align}
Note that the covariance matrices ${\bf Z}_1=\mathbb{E}[{\bf {z}}_1{\bf {z}}_1^*]$ and ${\bf Z}_2=\mathbb{E}[{\bf {z}}_2{\bf {z}}_2^*]$ are 
\begin{align}
{\bf Z}_1=  \sigma^2\left[%
\begin{array}{cc}
 1 & 0 \\
 0 & 1\!+\!\frac{1}{p_{1}^{\star}} \\
\end{array}%
\right]~~\rm{and}~~ {\bf Z}_2=  \sigma^2\left[%
\begin{array}{cc}
1 & 0 \\
 0 & 1\!+\!\frac{1}{p_{2}^{\star}} \\
\end{array}%
\right].
\end{align} 
Since we have used 3 channel uses, the achievable sum-rate of the two-user X-channel is 
\begin{align}
\sum_{k=1}^2\sum_{\ell=1}^2R_{k,\ell}=\frac{ \sum_{k=1}^2\log_2\left[ \det \left({\bf I} + P~{\bf {H}}_{k} {\bf {Z}}_k^{-1} {\bf {H}}_{k}^{*} \right)\right]}{3},
\end{align}
which completes the proof.
\endproof

To evaluate the performance of the proposed approach within a constant gap, it is instructive to compare our sum-rate result with an existing outer bound result in \cite{Cadambe_Jafar:09}, which is restated in the lemma below.
\begin{lemma}\label{lemma1}
The rate tuple $(R_{1,1}, R_{1,2}, R_{2,1}, R_{2,2})$ of the Gaussian two-user X-channel with the same set of channel coefficients $\{h_{i,j}[t]\}$ for $t\in\{1,4,9\}$ satisfies the following inequalities:
\begin{align}
R_{1,1} \!+\!R_{1,2}\!+\!R_{2,2} &\leq  \log_2\left\{1\!+\!\frac{\left(|h_{1,1}[t]|^2\!+\!|h_{1,2}[t]|^2\right)P}{\sigma^2}\right\}\nonumber \\&+\log_2\left\{1+\frac{ |h_{2,2}[t]|^2 P}{\sigma^2+ |h_{1,2}[t]|^2 P}\right\}, \\ 
R_{2,2} \!+\!R_{1,1}\!+\!R_{2,1}  &\leq  \log_2\left\{1+\frac{\left(|h_{2,2}[t]|^2\!+\!|h_{2,1}[t]|^2\right)P}{\sigma^2}\!\!\right\}\nonumber \\&+ \log_2\left\{1+\frac{ |h_{1,1}[t]|^2 P}{\sigma^2+ |h_{2,1}[t]|^2 P}\right\}, \\
R_{1,1} \!+\!R_{1,2}\!+\!R_{2,1} & \leq   \log_2\left\{1+\frac{\left(|h_{1,1}[t]|^2\!+\!|h_{1,2}[t]|^2\right)P}{\sigma^2}\right\} \nonumber \\&+\log_2\left\{1+\frac{ |h_{2,1}[t]|^2 P}{\sigma^2+ |h_{1,1}[t]|^2 P}\right\}, 
\end{align}
\begin{align}
R_{2,2} \!+\!R_{2,1}\!+\!R_{1,2} & \leq   \log_2\left\{\!\!1\!+\!\frac{\left(|h_{2,2}[t]|^2\!+\!|h_{2,1}[t]|^2\right)P}{\sigma^2}\right\}\nonumber \\&+\!\log_2\left\{1\!+\!\frac{ |h_{1,2}[t]|^2 P}{\sigma^2\!+\! |h_{2,2}[t]|^2 P}\right\}.
 \end{align}
\end{lemma}
\proof See \cite{Cadambe_Jafar:09}.
\endproof

\begin{corollary}
Consider channel coefficients which absolute values are one but with different phases, i.e., ${h_{i,k}[t]=e^{-jt\theta_{i,k}}}$ and ${\bf H}_1$ and ${\bf H}_2$ are orthogonal matrices, the sum-rate gap is bounded regardless of SNR as
\begin{align}
 \Delta R_{\Sigma}^X (\rm{SNR})\leq 2.39. \label{eq:Cov_prob_BS_coop}
\end{align}
\end{corollary}

\proof We prove the constant gap result using both Theorem \ref{Theorem3} and Lemma \ref{lemma1} for a particular set of channel values. Suppose the channel coefficients whose absolute values are one but with different phases, i.e., ${h_{i,k}[t]=e^{-jt\theta_{i,k}}}$ for $t\in\{1,4,9\}$. For this class of channels, from Lemma \ref{lemma1}, the sum-rate outer bound is given by 
\begin{align}
R_{1,1}+R_{2,1} \!+\!R_{1,2}\!+\!R_{2,2}&\leq \frac{4}{3}\log_2\left(1+2{\rm SNR}\right)\nonumber \\&+\frac{4}{3}\log_2\left(1+\frac{1}{1+\frac{1}{\rm SNR}}\right),\label{eq:outerex}
\end{align}
where ${\rm SNR}=\frac{P}{\sigma^2}$. Further, we assume that the phases of the channel coefficients are selected so that ${\bf H}_1$ and ${\bf H}_2$ are orthogonal matrices. Since $p_1^{\star}=p_2^{\star}=\frac{1}{2}$ from (\ref{eq:pwcont1}) and (\ref{eq:pwcont2}), the achievable sum rate of the proposed method is given by
\begin{align}
R_{1,1}+R_{2,1} \!+\!R_{1,2}\!+\!R_{2,2}&= \frac{2}{3}\log_2\left(1+2{\rm SNR}\right)\nonumber \\&+ \frac{2}{3}\log_2\left(1+\frac{2\rm{SNR}}{3}\right).\label{eq:innerex}
\end{align}
Therefore, the gap between the outer bound in (\ref{eq:outerex}) and the achievable rate in (\ref{eq:innerex}) is bounded as
\begin{align}
\Delta R_{\Sigma}^X (\rm{SNR}) &\!\leq  \frac{4}{3}\log_2\!\left(1\!+\!2{\rm SNR}\right)\!+\!\frac{4}{3}\log_2\!\left(\!1\!+\!\frac{1}{1\!+\!\frac{1}{\rm SNR}}\!\right)\nonumber \\&- \frac{2}{3}\!\left\{\!\log_2\left(1\!+\!2{\rm SNR}\right)\!+\!\log_2\left(\!1\!+\!\frac{2\rm{SNR}}{3}\!\right)\!\right\} \nonumber \\
&\leq \frac{2}{3}\! \log_2\!\left(\!\frac{1\!+\!2{\rm SNR}}{1\!+\!\frac{2}{3}\rm{SNR}}\!\right)\!+\!\frac{4}{3}\log_2\left(\!1\!+\!\frac{1}{1\!+\!\frac{1}{\rm SNR}}\!\right)\!\!.\label{eq:gap}
\end{align}
Since $\log_2\left(1+\frac{1}{1+\frac{1}{{\rm SNR}}}\right)\leq 1$ and $\log_2\left(\frac{1+2{\rm SNR}}{1+\frac{2}{3}{\rm SNR}}\right)\leq \log_2(3)$ for all ${\rm SNR}>0$, the gap further simplifies as
\begin{align}
 \Delta R_{\Sigma}^X (\rm{SNR})\leq \frac{2}{3}log_2(3)+\frac{4}{3}=2.39. \label{eq:gap2}
\end{align} This completes the proof.
\endproof
This corollary reveals that the proposed method achieves the sum-capacity of two-user X-channel within a constant gap for the entire SNR range 
for this particular class of channel coefficients, i.e., phase fading channels. This analysis should be carefully interpreted because it holds for the special sets of channel realizations.

\begin{figure}
\centering
\includegraphics[width=3.5in]{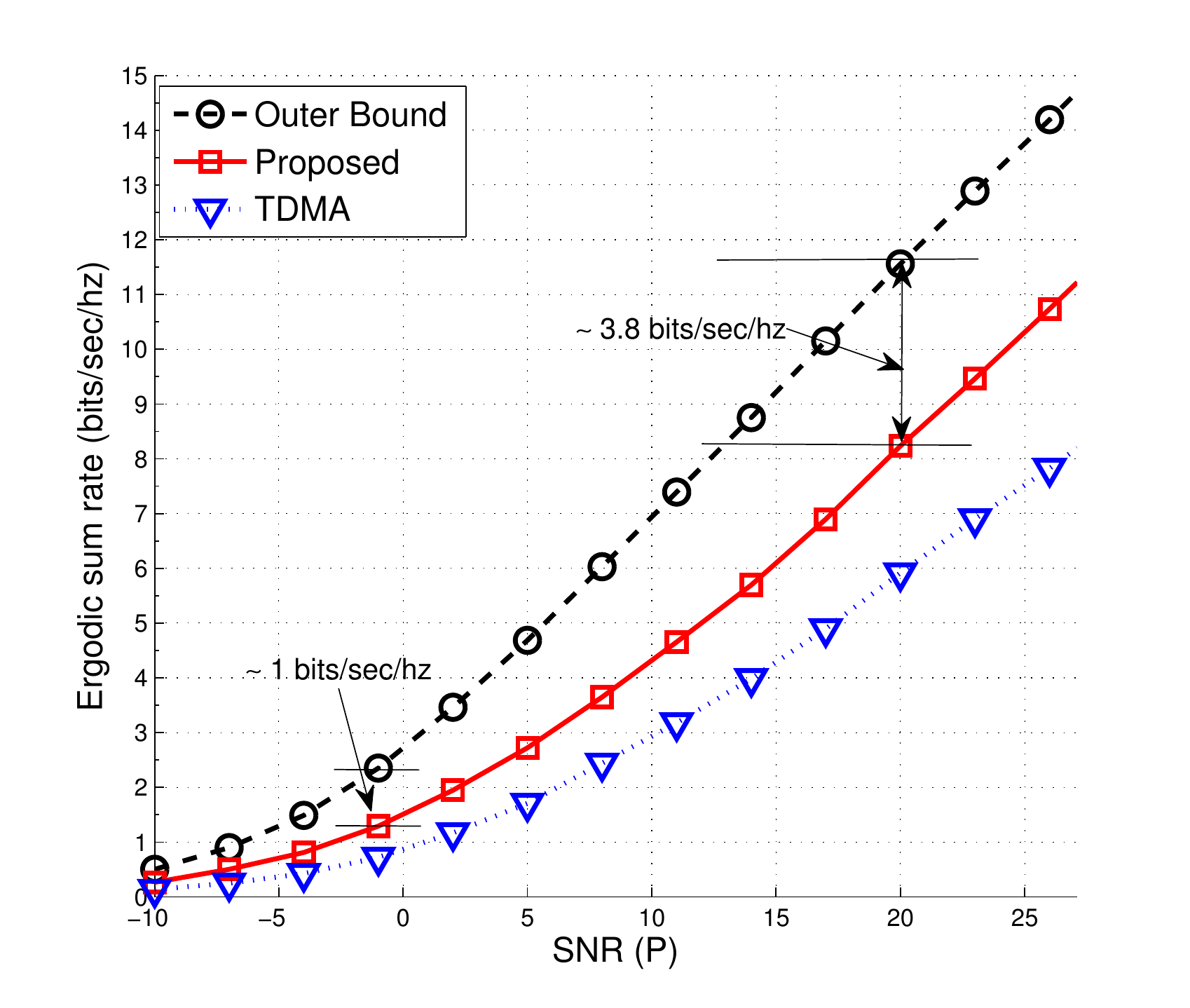}\vspace{-0.1cm}
\caption{The ergodic sum rate comparision bewteen the proposed interference alignment and the TDMA method for the two-user X-channel.} \label{fig:DoF}\vspace{-0.1cm}
\end{figure}

To provide a result for arbitrary channel realizations, the achievable ergodic rates of the two-user-X channel are compared with those obtained from the rate outer bound expression in \cite{Cadambe_Jafar:09} and TDMA transmission through simulations to demonstrate the superiority of the proposed method in the finite SNR regime. Fig. \ref{fig:DoF} illustrates the ergodic sum-rate obtained by the TDMA method, the proposed method, and outer bound expression in Lemma \ref{lemma1} when each channel is drawn from the complex Gaussian distribution, i.e., $\mathcal{CN}(0,1)$. One interesting observation is that the proposed interference alignment method always provides a better sum-rate than the TDMA method over the entire SNR regime, as the proposed method obtains the signal diversity gain from the repetition transmission method. Further, the proposed interference alignment achieves the ergodic sum-capacity of the two-user X-channel within a constant  number of bits $3.8$ bits/sec/hz over the entire range of SNR. In particular, in the low SNR regime, i.e., $\rm{SNR}<0 $ dB, the sum-capacity within one bit/sec/hz is achievable.


\section{Conclusion}

In this paper, we proposed a new interference management technique for a class of interference networks with local and moderately-delayed CSIT. With the proposed method, we characterized achievable trade-offs between the sum-DoF and CSI feedback in the interference networks with local CSIT. From the established trade-offs, we demonstrated the impact on how local and delayed CSIT affects the scale of network capacity in the interference networks. Further, by leveraging a known outer bound result, we showed that the proposed method achieves the sum-capacity of the two-user X-channel within a constant number of bits. 

Incorporating the effect of imperfect CSI and continuous block fading models would be desirable to refine and complete the analysis further. Another interesting direction for future study would be to investigate the effects of relays in interference networks with moderately-delayed CSIT by leveraging the idea of space-time physical layer network coding \cite{STPNC}.

\end{document}